\documentclass[twoside,leqno,twocolumn]{article}

\usepackage[letterpaper]{geometry}

\usepackage{siamproceedings}
\usepackage[T1]{fontenc}
\usepackage{amsmath,amssymb,amsfonts}
\usepackage{graphicx}
\usepackage{epstopdf}
\ifpdf
  \DeclareGraphicsExtensions{.eps,.pdf,.png,.jpg}
\else
  \DeclareGraphicsExtensions{.eps}
\fi

\usepackage{booktabs}
\usepackage{threeparttable}
\usepackage{enumitem}
\usepackage{multirow}
\usepackage{spverbatim}
\usepackage{url}
\usepackage{xurl}

\usepackage{xcolor}

\begin{document}

\newcommand\relatedversion{}

\title{\Large MasterSet: A Large-Scale Benchmark for Must-Cite Citation Recommendation in the AI/ML Literature}


\author{%
  Md Toyaha Rahman Ratul\thanks{Department of Computer Science, Northern Illinois University, DeKalb, IL
    (\email{mratul@niu.edu}, \email{zhanglei@niu.edu}).}
  \and Zhiqian Chen\thanks{Rochester Institute of Technology, Rochester, NY (\email{chen.zhiqian.work@gmail.com}).}
  \and Kaiqun Fu\thanks{Texas Christian University, Fort Worth, TX (\email{K.FU@tcu.edu}).}
  \and Taoran Ji\thanks{Texas A\&M University--Corpus Christi, Corpus Christi, TX (\email{taoran.ji@tamucc.edu}).}
  \and Lei Zhang\footnotemark[1]
}

\date{}

\maketitle

\fancyfoot[R]{\scriptsize{Copyright \textcopyright\ 2026 by SIAM\\
Unauthorized reproduction of this article is prohibited}}

\begin{abstract}
The explosive growth of AI and machine learning literature—with venues like NeurIPS and ICLR now accepting thousands of papers annually—has made comprehensive citation coverage increasingly difficult for researchers. While citation recommendation has been studied for over a decade, existing systems primarily focus on broad relevance rather than identifying the critical set of ``must-cite'' papers: direct experimental baselines, foundational methods, and core dependencies whose omission would misrepresent a contribution's novelty or undermine reproducibility. We introduce MasterSet, a large-scale benchmark specifically designed to evaluate must-cite recommendation in the AI/ML domain. MasterSet incorporates over 150,000 papers collected from official conference proceedings/websites of 15 leading venues, serving as a comprehensive candidate pool for retrieval. We annotate citations with a three-tier labeling scheme: (I) experimental baseline status, (II) core relevance (1–5 scale), and (III) intra-paper mention frequency. Our annotation pipeline leverages an LLM-based judge, validated by human experts on a stratified sample. The benchmark task requires retrieving must-cite papers from the candidate pool given only a query paper's title and abstract, evaluated by Recall@K. We establish baselines using sparse retrieval, dense scientific embeddings, and graph-based methods, demonstrating that must-cite retrieval remains a challenging open problem.
%
%
Code and data are available at \url{https://github.com/NIU-Graph-Intelligence/MasterSet}.
\end{abstract}

\section{Introduction.}
\label{sec:introduction}

The volume of published research in artificial intelligence and machine learning has grown dramatically over the past decade, with flagship venues now accepting thousands of papers annually---NeurIPS alone grew from 207 accepted papers in 2005 to 5,803 in 2025. For any researcher writing a new paper, this growth creates a formidable challenge: ensuring that all directly relevant prior work has been identified and cited. Missing a key citation is not merely an oversight---it can misrepresent the novelty of a contribution, undermine reproducibility by omitting comparison baselines, and in some cases constitute a form of academic misconduct when prior work is knowingly ignored. Yet no existing tool tells a researcher, given a paper or even just an idea, \emph{which specific papers they must cite}.

Citation recommendation systems have been studied for over a decade with the goal of addressing this problem, but their predominant framing---returning papers that are \emph{related} to a query in some general sense~\cite{9917887, farber2020citation, liang2023systematic}---is poorly matched to the practical need. A researcher finishing a paper does not need to discover every vaguely related work; they need to identify the small set of papers so central to their contribution---as baselines in their experiments, as the methods they directly extend, as the datasets they evaluate on---that omitting them would constitute an incomplete or misleading submission. We call these \textbf{must-cite} papers. 


Must-cite recommendation is a strictly harder task than general citation recommendation: the relevant set is small, the recall requirement is high, and missing even one is consequential. Our labels are derived from actual citation behavior---specifically, the subset that carries functional weight: papers explicitly used as experimental baselines, papers whose methods or datasets are directly built upon, and papers engaged with repeatedly throughout a manuscript. Like all citation benchmarks~\cite{lo2020s2orc, kang2018dataset}, our framework shares the unavoidable closed-world property that genuinely overlooked citations cannot appear as positive labels. What sets MasterSet apart is the signal we extract: rather than treating every cited paper as equally relevant, we identify the functionally indispensable subset---the only operationalisation of must-cite status that is tractable at dataset scale.

To enable systematic progress on this task, we introduce \textbf{MasterSet}, the first large-scale benchmark specifically designed to evaluate must-cite recommendation in the AI and ML literature. MasterSet is built on a collection of \textbf{153,373} papers\footnote{ This count reflects the current collection snapshot at the time of submission and may increase as new conference years are added. Annotations are added continuously.} from \textbf{15} peer-reviewed venues spanning core ML, general AI, computer vision, natural language processing, and probabilistic methods, collected from official proceedings/websites to yield an authoritative, exact paper count free of preprint conflation. To our knowledge this is the most comprehensive openly available collection of AI and ML papers from the deep learning era, covering all nine top-ranked venues in Google Scholar's AI and Computer Vision h5-index rankings. We annotate every citation instance with a three-tier scheme capturing experimental baseline status (Type~I), core relevance on a 1--5 scale (Type~II), and intra-paper mention frequency (Type~III), executed at scale with an LLM judge and validated against human labels on a stratified sample of 510 instances. The benchmark task is: given the title and abstract of a query paper, retrieve from the candidate pool set, those papers that are must-cite for the query. Evaluation uses Recall@$K$ as the primary metric, reflecting the asymmetric cost of missing a must-cite paper.

Our main contributions are:
\begin{enumerate}[leftmargin=*]
  \item \textbf{Task definition.} We formally define the must-cite recommendation task, distinguishing it from general and local citation recommendation, and argue for its importance as a tool for citation integrity.

  \item \textbf{Dataset.} We construct MasterSet by writing venue-specific crawlers for all 15 official conference websites, yielding a verified, exact paper count free of preprint conflation. Unlike aggregator services such as Semantic Scholar and S2ORC~\cite{lo2020s2orc}, which report inconsistent counts and may conflate preprint versions with final published papers, our official-proceedings collection provides exact, authoritative totals. Unlike domain-specific datasets such as ACL-ARC~\cite{bird2008acl}, which is limited to the NLP/CL literature, MasterSet spans five subfields of AI and ML across 15 venues.

  \item \textbf{Annotation framework and execution.} We introduce a three-tier labeling scheme providing a principled, multi-faceted operationalisation of must-cite status, reusable independently of the specific LLM judge employed. We execute it across the full dataset with an LLM judge, validate labels through inter-run consistency and a human validation study, and report annotation cost transparently.

  \item \textbf{Baselines.} We evaluate sparse, dense, and graph-based retrieval methods, establishing the first performance reference for must-cite retrieval and showing the task remains an open challenge.
\end{enumerate}

\section{Related Work.}
\label{sec:related}
\leavevmode\par

\textbf{Citation Recommendation.} Citation recommendation has been studied in two settings. \emph{Local} recommendation predicts which paper fills a specific citation placeholder given surrounding context~\cite{ebesu2017neural, gu2022local, goyal2024symtax}; because it presupposes the author has already identified a gap, it cannot surface papers the author does not know they should cite. \emph{Global} recommendation takes the full manuscript or its metadata and suggests papers for the bibliography~\cite{farber2020citation, liang2023systematic}. Systems range from sparse retrieval (BM25~\cite{robertson1995okapi}) to dense scientific embeddings (SPECTER~\cite{cohan2020specter}, SPECTER2~\cite{Singh2022SciRepEvalAM}, SciNCL~\cite{Ostendorff2022scincl}) and structure-aware methods that incorporate citation graphs~\cite{ali2024glamor, kammari2024relevant}. A fundamental limitation of all these systems is that they optimise for \emph{topical relevance}: any cited paper is a positive label, with no distinction between a core experimental baseline and tangential background. The work most closely related to MasterSet is CitationR~\cite{long2024recommending}, which uses reviewer-identified missed citations as gold labels. This is a valuable signal, but such labels are expensive to collect, limited in scale, and specific to the peer-review context; MasterSet operationalises must-cite status at dataset scale through three complementary criteria (Section~\ref{sec:annotation}).

\textbf{Citation Intent Classification.}~Citation intent classification assigns functional labels to citation instances, distinguishing background references from methods being extended or compared against. Key resources include ACL-ARC~\cite{bird2008acl}, the citation frame scheme of~\cite{jurgens2018measuring}, and SciCite~\cite{cohan2019structural}. Kunnath et al.~\cite{kunnath2023prompting} find that zero-shot LLMs struggle to distinguish background from comparison citations without few-shot anchoring---an insight we adopt directly in our Type~I annotation prompt. Unlike prior work where intent labels are the end goal, in MasterSet they serve as one component of a broader must-cite definition that also includes relevance scoring and mention frequency.

\textbf{LLM-as-Judge for Annotation.}~Zheng et al.~\cite{zheng2023judging} show that GPT-4 judgements correlate well with human preferences but identify failure modes including position bias and high parse error rates for open-ended outputs. We apply their key insight directly: our prompts require a single-digit response, reducing parse failures to essentially zero. Unlike MT-Bench, where LLM judgements are compared against human preference, our setting uses them as \emph{ground truth labels}, placing a higher reliability bar that motivates our human validation study (Section~\ref{sec:validation}).

\textbf{Scholarly Datasets and Benchmarks.}~Existing large-scale scholarly corpora---S2ORC~\cite{lo2020s2orc}, PeerRead~\cite{kang2018dataset}, OpenCitations~\cite{peroni2020opencitations}, and citation-context datasets~\cite{ostendorff2022neighborhood}---record which papers were cited but do not distinguish must-cite from optional citations. MasterSet differs in three respects: it is restricted to AI and ML, providing a domain-focused candidate pool; it is sourced exclusively from official proceedings/websites, yielding exact, verified paper counts; and most importantly, it provides must-cite labels---the property that makes it a benchmark rather than merely a corpus.

\section{Dataset Construction.}
\label{sec:dataset}

MasterSet is constructed in two stages: paper collection (Section~\ref{sec:venues} and \ref{sec:collection}) and citation context extraction (Section~\ref{sec:extraction}). Annotation methodology is described in Section~\ref{sec:annotation}.

\subsection{Venue Selection.}
\label{sec:venues}

We collect papers from 15 peer-reviewed venues spanning the core subfields of artificial intelligence and machine learning. The selection is organized around five groups: (1)~\emph{core ML}: NeurIPS, ICML, ICLR; (2)~\emph{general AI}: AAAI, IJCAI; (3)~\emph{computer vision}: CVPR, ICCV, ECCV; (4)~\emph{natural language processing}: ACL, EMNLP, NAACL; and (5)~\emph{theory and probabilistic methods}: COLT, UAI, AISTATS. We also include JMLR, the only journal (as opposed to conference proceedings) in our collection. The temporal scope of the collection begins with the deep learning era. The majority of venues are collected from 2012 or 2013 onward, a period widely recognized as the inflection point at which deep neural networks became the dominant paradigm in AI and ML research~\cite{krizhevsky2012imagenet}.

For venues with longer histories---NeurIPS (from 2000), JMLR (from 2000), AISTATS (from 2009), AAAI (from 2010), COLT (from 2011)---earlier proceedings are included where they are publicly accessible, though comprehensive coverage is not guaranteed prior to 2012. Detailed discussion on venue representativeness is provided in Appendix~\ref{app:ethics}. 

\subsection{Data Collection.}
\label{sec:collection}

All papers are collected directly from the official websites of each venue (e.g., \url{papers.nips.cc}, \url{proceedings.mlr.press}, \url{openreview.net}, \url{openaccess.thecvf.com}). This design choice is deliberate. Aggregator services such as Google Scholar and Semantic Scholar have been observed to return inconsistent paper counts for specific conference-year pairs, and may conflate preprint versions with final published versions\footnote{See Appendix~\ref{app:s2orc} for per-venue discrepancy statistics.}. By contrast, the official website pages constitute the authoritative record of accepted papers, and the paper counts obtained from them are exact.

For each paper we collect: title, authors, abstract, and PDF. All metadata is scraped at the paper level from the official page; PDFs are downloaded from the same source for internal use in the citation extraction pipeline (Section~\ref{sec:extraction}) and are not redistributed.

\paragraph{Main conference filtering.}
Several venues intermix main-conference papers with co-located workshops, tutorials, shared tasks, and system demonstrations within the same proceedings index. We retain only main-conference full papers and exclude all ancillary tracks. For venues whose proceedings structure makes this distinction unambiguous (e.g., NeurIPS, ICML, CVPR), no additional filtering is required. For venues with more complex structures---specifically AAAI, IJCAI, ACL, EMNLP, and NAACL---we apply a two-stage filtering procedure. First, we enumerate all track or volume names listed in the proceedings index for a given year. Second, we prompt a large language model (Gemini 2.5 Flash) with the conference name, year, and the full list of track names, and instruct it to identify which tracks constitute the main conference program. The model's classifications are cached and can be corrected manually if needed. For AAAI, which bundles multiple tracks across several issues of its open-access journal, filtering is applied at both the issue level and the section level within each issue. This procedure ensures that inflated paper counts from non-main-track content do not contaminate the dataset.

\subsection{Dataset Statistics.}
\label{sec:statistics}

Table~\ref{tab:dataset} reports the number of collected papers per venue, along with the year range and total count. As of the time of writing, the dataset comprises 153,373 papers across 15 venues and 198 conference-years. The dataset is the product of an ongoing collection effort; paper counts will increase as future proceedings are published and incorporated.

\begin{table}[t]
\centering
\caption{ Overview of collected papers by venue. Year ranges reflect the scope of the current collection snapshot; earlier proceedings exist for some venues but are not fully covered. ECCV and ICCV are biennial and appear in alternate years only. NAACL was not held every year throughout the collection period. }
\label{tab:dataset}
\footnotesize
\begin{threeparttable}
\begin{tabular}{llrr}
\toprule
\textbf{Venue} & \textbf{Type} & \textbf{Years} & \textbf{Papers} \\
\midrule
NeurIPS & Conference & 2000--2025$^\ddagger$ & 28,829 \\
CVPR    & Conference & 2013--2025 & 18,446 \\
AAAI    & Conference & 2010--2025 & 17,755 \\
ICLR    & Conference & 2013--2026$^\ddagger$ & 16,840 \\
ICML    & Conference & 2013--2025 & 14,281 \\
ICCV    & Conference & 2013--2025$^\dagger$ & 9,145 \\
ACL     & Conference & 2017--2025 & 9,693 \\
EMNLP   & Conference & 2017--2025 & 9,508 \\
ECCV    & Conference & 2018--2024$^\dagger$ & 6,166 \\
IJCAI   & Conference & 2017--2025 & 7,042 \\
AISTATS & Conference & 2009--2025 & 4,697 \\
JMLR    & Journal    & 2000--2025 & 3,932 \\
NAACL   & Conference & 2013--2025 & 3,716 \\
UAI     & Conference & 2015--2025 & 1,740 \\
COLT    & Conference & 2011--2025 & 1,583 \\
\midrule
\textbf{Total} & & & \textbf{153,373} \\
\bottomrule
\end{tabular}
\begin{tablenotes}
\small
\item[$\dagger$] Biennial conference; odd or even years only.
\item[$\ddagger$] Most recent year collected but annotation not yet
complete.
\end{tablenotes}
\end{threeparttable}
\end{table}

The distribution is skewed toward the most established and highest-volume venues: NeurIPS, CVPR, AAAI, ICLR, and ICML together account for roughly 63\% of the collection, while UAI and COLT---the smallest venues---together contribute fewer than 3,400 papers. This imbalance is inherent to the field rather than an artifact of our collection strategy; it reflects the differing scales of these communities. 

\subsection{Citation Extraction.}
\label{sec:extraction}
For each paper we extract its full reference list and in-text citation contexts using a two-tier pipeline. Our primary tool is Nougat~\cite{blecher2023nougat}, a transformer-based PDF-to-markup model that recovers logical document structure---section headings, equations, and reference lists---well-suited to the IMRaD format common in AI  papers. From Nougat's output we parse each reference entry and locate every in-text occurrence, recording for each a \textbf{three-sentence context window}---comprising the citing sentence together with one sentence of preceding and following context, sufficient to determine experimental intent without introducing noise from distant text---together with the \textbf{extracted section heading} (e.g., \emph{4~Experiments}). 
The section label is critical downstream: Type~I annotation uses it to distinguish experimental comparisons from background citations (Section~\ref{sec:annotation}).

Nougat exhibits three systematic failure modes---reference reordering, OCR digit confusion, and alpha-tag/numeric mismatch---that silently corrupt citation linking (details in Appendix~\ref{app:failures}). Papers triggering any of these cases are processed instead with GROBID~\cite{grobid}, which operates directly on the PDF rendering stream and produces TEI~XML with explicitly annotated citation spans, bypassing all three failure modes. Both paths emit the same JSON schema, ensuring full downstream compatibility (schema specification in Appendix~\ref{app:schema}). Coverage statistics are reported in Appendix~\ref{app:pipeline}.
\section{Annotation Methodology.}
\label{sec:annotation}

MasterSet labels are produced through a three-tier scheme. Types~I and~II are annotated at scale using an LLM judge; Type~III is computed directly from citation frequency.

\subsection{Three-Tier Labeling Scheme.}
\label{sec:tiers}

\paragraph{Type~I: Experimental Baseline (Binary).}
A binary label $y^{\mathrm{I}} \in \{0,1\}$ indicating whether the cited work is explicitly used as a direct comparison baseline in the current paper's experiments or results section, the source of a benchmark, dataset, or evaluation task used in the paper, or a core task or method the paper directly builds on. This is the most objective form of must-cite status: a paper that claims to outperform a method without citing it makes an unverifiable claim. Full annotation criteria are given in Appendix~\ref{app:tier_criteria}.

\paragraph{Type~II: Core Relevance (1--5 Scale).}
A score $y^{\mathrm{II}} \in \{1,\ldots,5\}$ assessing how central the cited work is to the query paper's core task and method, from general background (1) to direct task-and-method overlap (5). Papers scoring 4 or 5 are considered must-cite under Type~II. Two hard rules apply regardless of context: a citation to the primary dataset or benchmark receives at least 4; a citation to a directly extended method receives at least 4. Full rubric and anchor examples are in Appendix~\ref{app:tier_criteria}.

\paragraph{Type~III: Intra-Paper Mention Frequency.}
A paper $p$ receives a Type~III must-cite flag for query paper $q$ if $p$ is mentioned at least $N$ times within $q$. Repeated engagement across motivation, methodology, experiments, and discussion signals a core dependency rather than incidental background. This tier requires no LLM annotation; counts are computed directly from the extracted citation graph. The threshold $N = \textbf{3}$ is determined by inspecting the within-paper mention frequency distribution (Appendix~\ref{app:tier_criteria}).

\subsection{LLM Annotation Pipeline.}
\label{sec:llm_pipeline}

Types~I and~II are annotated using Gemini~2.5 Flash~\cite{gemini25flash} as an LLM judge, applied independently to each citation instance. Each instance consists of the query paper title, the cited paper's bibliographic information, and all extracted citation contexts with their section labels (Section~\ref{sec:extraction}). Prompts follow the guidelines of~\cite{kunnath2023prompting, zheng2023judging}: few-shot anchoring, an explicit scope guard requiring any qualifying comparison to occur \emph{in the current paper}, and structured single-digit output to eliminate parse failures. The model is run in greedy decoding mode (temperature~$= 0$) to maximise determinism. Both tasks are run in separate passes to avoid order effects. Full prompts are reproduced in Appendix~\ref{app:prompts}. Details regarding model selection, API specifications, and the total cost for annotating all citation instances are reported in Appendix~\ref{app:cost}. 



\begin{figure*}[t]
\centering
\includegraphics[width=\textwidth]{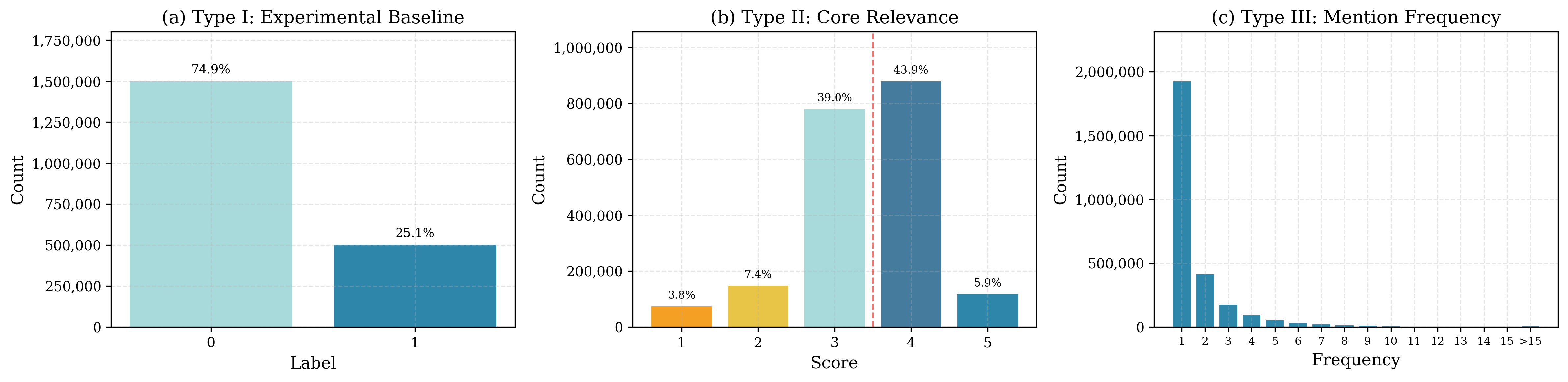}
\caption{Label distributions across the three must-cite tiers. (a)~Type~I is a binary baseline label with a 1:3 positive-to-negative ratio. (b)~Type~II is a 1--5 relevance score; the dashed line marks the must-cite threshold ($y^{\mathrm{II}} \geq 4$). (c)~Type~III is intra-paper mention frequency, truncated at 15 for display; the distribution is heavily right-skewed, with the vast majority of references mentioned only once or twice.}
\label{fig:label_dist}
\end{figure*}

\subsection{Human Validation Study.}
\label{sec:validation}

We conduct a human validation study with six active AI/ML researchers to verify label reliability against expert judgment. The annotation sample comprises 510 citation instances drawn from the full MasterSet corpus via stratified sampling across the 15 venues and 2000--2025 year range. Each instance is independently labeled by between two and six annotators. Every instance is judged on Q1 (Type~I, binary) and Q2 (Type~II, 1--5) with citation contexts highlighted in-line. Full annotation instructions, the interface, and the sampling protocol are detailed in Appendix~\ref{app:protocol}.

Inter-annotator agreement is reported as Krippendorff's~$\alpha$~\cite{krippendorff2011computing} (nominal for Type~I, ordinal for Type~II), which supports designs with a variable number of raters per item. Both tasks fall in the moderate range ($\alpha = 0.461$, 0.492), consistent with prior citation-annotation work on similarly subjective labels~\cite{cohan2019structural, jurgens2018measuring}.

\paragraph{LLM--human agreement.}
We compare Gemini~2.5 Flash predictions against the human-annotated labels using accuracy for Type~I and quadratic-weighted Cohen's~$\kappa$ for Type~II. The Type~II weighted~$\kappa$ uses the same quadratic weighting as the pairwise human agreement computed above, so the two Type~II numbers are on the same scale. On both tasks, the LLM--human agreement (Table~\ref{tab:agreement}) falls within the range of pairwise disagreement observed among the human annotators themselves---Gemini disagrees with the human labels no more than expert annotators disagree with each other, which we take as the reliability criterion justifying the LLM judge at corpus scale.

\begin{table}[t]
\centering
\caption{Agreement results on the human validation set ($n=510$
instances, 2--6 annotators each).}
\label{tab:agreement}
\small
\begin{threeparttable}
\begin{tabular}{lcc}
\toprule
 & \textbf{Inter-annotator} & \textbf{LLM--human} \\
\midrule
Type~I  & 0.461$^a$ & 0.816$^b$ \\
Type~II & 0.492$^a$ & 0.355$^c$ \\
\bottomrule
\end{tabular}
\begin{tablenotes}
\small
\item[$a$] Krippendorff's $\alpha$ (nominal for Type~I, ordinal for Type~II).
\item[$b$] Accuracy.
\item[$c$] Quadratic-weighted Cohen's $\kappa$.
\end{tablenotes}
\end{threeparttable}
\end{table}
\section{Ethical Considerations.}
\label{sec:ethics}

We discuss data licensing, coverage limitations, risks of LLM-generated ground truth, and dual-use considerations in Appendix~\ref{app:ethics}. In brief: all papers are collected from open-access official proceedings/websites without circumventing any access controls; we release only metadata, citation graphs, and LLM-generated labels, not the PDFs themselves; LLM-generated labels carry inherent systematic bias risks that we partially mitigate through prompt design and quantify through the human validation study (Section~\ref{sec:validation}); and the benchmark is intended as a tool for scholarly diligence, not a substitute for researcher judgement.
\section{Benchmark Task.}
\label{sec:benchmark}

\subsection{Task Definition.}
\label{sec:task_def}

The must-cite retrieval task is defined as follows. Given the title and abstract of a query paper~$q$, retrieve from the MasterSet-CoreML-v1 candidate pool (Section~\ref{sec:eval_subset}) a ranked list of papers that are must-cite for~$q$. A paper~$p$ is must-cite for~$q$ under our framework if it satisfies at least one of the following criteria derived from Section~\ref{sec:annotation}: (i)~$y^{\mathrm{I}}(p,q) = 1$ (Type~I: experimental baseline), (ii)~$y^{\mathrm{II}}(p,q) \geq 4$ (Type~II: substantially or core-relevant), or (iii)~$p$ is mentioned at least $N$ times within~$q$ (Type~III).

The input signal is deliberately restricted to title and abstract. This reflects the realistic scenario of a researcher with an early draft or a well-developed idea who wants to audit citation coverage before writing is complete, without requiring access to the full manuscript text. Evaluation uses Recall@$K$ as the primary metric, reflecting the asymmetric cost of missing a must-cite paper. We also report MAP, MRR, nDCG@$K$, and HR@$K$ to characterise ranking quality at different depths.

\subsection{Evaluation Subset.}
\label{sec:eval_subset}

MasterSet is released as a living resource: the full collection of 153,373 papers across all 15 venues, together with $\approx 91\%$ completed annotations, is made publicly available and updated as new proceedings are published and labeled. 
The labeled papers set forms the basis of all benchmark experiments and the label distribution reported in Figure~\ref{fig:label_dist}.
Within this release we define \textbf{MasterSet-CoreML-v1}, a versioned benchmark split that enables direct, reproducible comparison across systems. 


\paragraph{Core set construction.}
The benchmark candidate pool is derived from the already labelled papers in two stages. First, we extract the \emph{core set}: all papers from the three target venues (NeurIPS, ICML, ICLR), together with any paper from the remaining twelve venues that is directly cited by at least one target-venue paper. Papers from the other twelve venues that are never cited by a NeurIPS, ICML, or ICLR paper are excluded from the pool, since they cannot be a must-cite target for any query in our benchmark. Second, we apply the temporal split below to partition the core set into a training/candidate pool and an evaluation set.

\paragraph{Candidate pool.}
The retrieval pool for all experiments consists of \textbf{67,761} core-set papers published between 2018 and 2024 inclusive. This pool serves as both the candidate set over which all retrieval methods rank, and the training set available for supervised fine-tuning. Papers published before 2018 are retained in the labelled corpus but not in the pool (see \emph{Temporal split} below).

\paragraph{Query venue selection.}
Query papers are drawn exclusively from NeurIPS, ICML, and ICLR---the three highest-volume venues in the collection and the primary publication venues of the core ML community, ensuring that must-cite candidates span all subfields in the pool rather than being domain-restricted. The remaining twelve venues contribute to the candidate pool only through the citation-linking step.

\paragraph{Temporal split.}
We partition the core set by publication year:
\begin{itemize}[leftmargin=*]
  \item \textbf{Train} (2018--2024): the 67,761 core-set papers from this period, available for supervised fine-tuning of retrieval models and serving as the fixed candidate pool for ranking.
  \item \textbf{Evaluation} (2025): 7,028 papers from NeurIPS, ICML, and ICLR published in 2025, constituting the fixed evaluation query set for benchmark reporting.
\end{itemize}
The temporal cut ensures that all must-cite candidates for test queries are already present in the pool, and that the test set reflects the realistic scenario of recommending citations for recently submitted papers. Each query paper must have at least one must-cite paper in the pool under the combined criterion of Section~\ref{sec:task_def}; papers not satisfying this condition are excluded as trivially unevaluable.

\paragraph{Versioning.}
The \textit{v1} designation fixes the pool snapshot, split boundaries, and annotation version used in this paper. Future releases extending the query venues, time range, or annotation scheme will be issued as \textit{v2}, \textit{v3}, and so on, preserving backward comparability with reported results.
\section{Experiments.}
\label{sec:results}

\subsection{Experimental Setup.}
\label{sec:setup}

All methods receive the title and abstract of the query paper as input and retrieve from the 67,761-paper candidate pool. We evaluate on the MasterSet-CoreML-v1 split defined in Section~\ref{sec:eval_subset}. Evaluation metrics are MAP, MRR, nDCG@$K$ ($K \in \{10,20,30\}$), Recall@$K$ ($K \in \{50, 100\}$), and HR@$K$ ($K \in \{10, 20\}$); full metric definitions are given in Appendix~\ref{app:metrics}.

\subsection{Baseline Methods.}
\label{sec:baselines}

We evaluate three categories of baselines: (1)~sparse retrieval, (2)~semantic retrieval and representation models, and (3)~structure-aware and graph reasoning models. All methods receive only the title and abstract of the query paper. Detailed descriptions of each method are in Appendix~\ref{app:baselines}.

\paragraph{Sparse retrieval.}
\textbf{BM25}~\cite{robertson1995okapi} ranks candidates by keyword overlap with length-normalised term weighting.

\paragraph{Semantic retrieval.}
\textbf{SciBERT}~\cite{beltagy2019scibert} is evaluated in two configurations: frozen (vanilla), and fine-tuned with NT-Xent contrastive loss (SciBERT-NTX). \textbf{SPECTER}~\cite{cohan2020specter}, \textbf{SPECTER2}~\cite{Singh2022SciRepEvalAM}, and \textbf{SciNCL}~\cite{Ostendorff2022scincl} are citation-informed pre-trained embeddings supporting cold-start inference from title and abstract. \textbf{ColBERT}~\cite{khattab2020colbert} uses late interaction with MaxSim scoring, evaluated vanilla and fine-tuned (ColBERT-NTX). \textbf{HAtten-RR}~\cite{gu2022local} is a two-stage prefetch-and-rerank pipeline with hierarchical attention.

\paragraph{Graph-based retrieval.}
\textbf{KTR}~\cite{wu2024supporting}, \textbf{AR-GNN}~\cite{kammari2024relevant}, and \textbf{LitFM}~\cite{zhang2025litfm} incorporate citation graph structure via topic-level reasoning paths, heterogeneous GNNs, and GNN-augmented encoders respectively.

\subsection{Main Results.}
\label{sec:main_results}

\begin{table*}[ht]
\centering
\caption{\textbf{Type~I} (Experimental Baseline) Retrieval performance on MasterSet-CoreML-v1 (title + abstract input). Best result per column in \textbf{bold}.
$\diamond$~Unsupervised, no learned parameters.
$\dagger$~Pretrained on external data; used zero-shot on MasterSet.
$\ddagger$~Pretrained on external data, then fine-tuned on MasterSet.
$\star$~Implemented from scratch and trained on MasterSet.}
\label{tab:type1_results}
\footnotesize
\begin{tabular}{l c c c c c c c c c}
\toprule
\textbf{Method} & \textbf{MAP} & \textbf{MRR} & \textbf{nD@10} & \textbf{nD@20} & \textbf{nD@30} & \textbf{R@50} & \textbf{R@100} & \textbf{HR@10} & \textbf{HR@20} \\
\midrule
BM25$^{\diamond}$          & \textbf{0.0929} & \textbf{0.2800} & \textbf{0.1396} & \textbf{0.1529} & \textbf{0.1621} & 0.2487 & 0.3006 & \textbf{0.4468} & \textbf{0.5335} \\
SciBERT$^{\dagger}$        & 0.0045 & 0.0175 & 0.0072 & 0.0083 & 0.0092 & 0.0188 & 0.0264 & 0.0337 & 0.0488 \\
SciBERT-NTX$^{\ddagger}$   & 0.0916 & 0.2405 & 0.1284 & 0.1470 & 0.1618 & \textbf{0.2963} & \textbf{0.3832} & 0.4393 & \textbf{0.5335} \\
SPECTER$^{\dagger}$        & 0.0776 & 0.2255 & 0.1144 & 0.1279 & 0.1378 & 0.2323 & 0.2895 & 0.3921 & 0.4855 \\
SPECTER2$^{\dagger}$       & 0.0905 & 0.2517 & 0.1324 & 0.1469 & 0.1581 & 0.2624 & 0.3258 & 0.4359 & 0.5235 \\
SciNCL$^{\dagger}$         & 0.0858 & 0.2383 & 0.1248 & 0.1407 & 0.1522 & 0.2565 & 0.3223 & 0.4174 & 0.5143 \\
ColBERT$^{\dagger}$        & 0.0646 & 0.1962 & 0.0987 & 0.1105 & 0.1190 & 0.2000 & 0.2497 & 0.3505 & 0.4360 \\
ColBERT-NTX$^{\ddagger}$   & 0.0888 & 0.2460 & 0.1289 & 0.1446 & 0.1571 & 0.2677 & 0.3416 & 0.4328 & 0.5215 \\
HAtten-RR$^{\star}$        & 0.0004 & 0.0012 & 0.0005 & 0.0007 & 0.0008 & 0.0017 & 0.0031 & 0.0023 & 0.0049 \\
KTR$^{\star}$              & 0.0025 & 0.0094 & 0.0039 & 0.0046 & 0.0052 & 0.0111 & 0.0172 & 0.0185 & 0.0283 \\
AR-GNN$^{\star}$           & 0.0049 & 0.0202 & 0.0070 & 0.0085 & 0.0108 & 0.0294 & 0.0573 & 0.0488 & 0.0704 \\
LitFM$^{\star}$            & 0.0406 & 0.1516 & 0.0658 & 0.0755 & 0.0835 & 0.1580 & 0.2220 & 0.2970 & 0.3884 \\
\bottomrule
\end{tabular}%

\end{table*}

\begin{table*}[ht]
\centering
\caption{\textbf{Type~II} (Core Relevance, score $\geq 4$) Retrieval performance on MasterSet-CoreML-v1 (title + abstract input). Notation and baseline setup are identical to Table~\ref{tab:type1_results}}
\label{tab:type2_results}
\footnotesize
\begin{tabular}{l c c c c c c c c c}
\toprule
\textbf{Method} & \textbf{MAP} & \textbf{MRR} & \textbf{nD@10} & \textbf{nD@20} & \textbf{nD@30} & \textbf{R@50} & \textbf{R@100} & \textbf{HR@10} & \textbf{HR@20} \\
\midrule
BM25$^{\diamond}$          & 0.0917 & \textbf{0.3913} & 0.1750 & 0.1719 & 0.1805 & 0.2350 & 0.2922 & 0.6148 & 0.6971 \\
SciBERT$^{\dagger}$        & 0.0045 & 0.0324 & 0.0101 & 0.0105 & 0.0113 & 0.0185 & 0.0270 & 0.0625 & 0.0903 \\
SciBERT-NTX$^{\ddagger}$   & \textbf{0.1181} & 0.3882 & \textbf{0.1929} & \textbf{0.1995} & \textbf{0.2145} & \textbf{0.3163} & \textbf{0.4102} & \textbf{0.6454} & \textbf{0.7469} \\
SPECTER$^{\dagger}$        & 0.0809 & 0.3340 & 0.1495 & 0.1493 & 0.1581 & 0.2185 & 0.2785 & 0.5572 & 0.6498 \\
SPECTER2$^{\dagger}$       & 0.0970 & 0.3711 & 0.1749 & 0.1742 & 0.1849 & 0.2536 & 0.3216 & 0.6065 & 0.6955 \\
SciNCL$^{\dagger}$         & 0.0930 & 0.3543 & 0.1658 & 0.1672 & 0.1776 & 0.2475 & 0.3169 & 0.5888 & 0.6858 \\
ColBERT$^{\dagger}$        & 0.0697 & 0.2953 & 0.1317 & 0.1317 & 0.1401 & 0.1938 & 0.2486 & 0.5068 & 0.5993 \\
ColBERT-NTX$^{\ddagger}$   & 0.1038 & 0.3725 & 0.1790 & 0.1811 & 0.1935 & 0.2755 & 0.3543 & 0.6134 & 0.7051 \\
HAtten-RR$^{\star}$        & 0.0001 & 0.0020 & 0.0003 & 0.0004 & 0.0005 & 0.0013 & 0.0026 & 0.0032 & 0.0060 \\
KTR$^{\star}$              & 0.0026 & 0.0195 & 0.0056 & 0.0064 & 0.0071 & 0.0130 & 0.0193 & 0.0392 & 0.0626 \\
AR-GNN$^{\star}$           & 0.0051 & 0.0281 & 0.0092 & 0.0097 & 0.0114 & 0.0245 & 0.0451 & 0.0700 & 0.1003 \\
LitFM$^{\star}$            & 0.0571 & 0.2595 & 0.1120 & 0.1119 & 0.1204 & 0.1824 & 0.2520 & 0.4944 & 0.5912 \\
\bottomrule
\end{tabular}%
\end{table*}

\begin{table*}[ht]
\centering
\caption{\textbf{Type~III} (Mention Frequency, $\geq 3$ mentions) Retrieval performance on MasterSet-CoreML-v1 (title + abstract input). Notation and baseline setup are identical to Table~\ref{tab:type1_results}.}
\label{tab:type3_results}
\footnotesize
\begin{tabular}{l c c c c c c c c c}
\toprule
\textbf{Method} & \textbf{MAP} & \textbf{MRR} & \textbf{nD@10} & \textbf{nD@20} & \textbf{nD@30} & \textbf{R@50} & \textbf{R@100} & \textbf{HR@10} & \textbf{HR@20} \\
\midrule
BM25$^{\diamond}$          & \textbf{0.1405} & \textbf{0.3512} & \textbf{0.1967} & \textbf{0.2166} & \textbf{0.2288} & 0.3466 & 0.4139 & \textbf{0.5523} & \textbf{0.6346} \\
SciBERT$^{\dagger}$        & 0.0066 & 0.0225 & 0.0099 & 0.0116 & 0.0127 & 0.0255 & 0.0361 & 0.0414 & 0.0598 \\
SciBERT-NTX$^{\ddagger}$   & 0.1298 & 0.2931 & 0.1741 & 0.2023 & 0.2193 & \textbf{0.3970} & \textbf{0.4932} & 0.5227 & 0.6329 \\
SPECTER$^{\dagger}$        & 0.1041 & 0.2661 & 0.1457 & 0.1656 & 0.1783 & 0.3055 & 0.3727 & 0.4544 & 0.5515 \\
SPECTER2$^{\dagger}$       & 0.1273 & 0.3069 & 0.1758 & 0.1971 & 0.2112 & 0.3482 & 0.4221 & 0.5134 & 0.6063 \\
SciNCL$^{\dagger}$         & 0.1193 & 0.2853 & 0.1636 & 0.1859 & 0.2001 & 0.3384 & 0.4163 & 0.4909 & 0.5884 \\
ColBERT$^{\dagger}$        & 0.0918 & 0.2320 & 0.1299 & 0.1464 & 0.1576 & 0.2686 & 0.3317 & 0.4145 & 0.4978 \\
ColBERT-NTX$^{\ddagger}$   & 0.1349 & 0.3165 & 0.1843 & 0.2082 & 0.2237 & 0.3732 & 0.4579 & 0.5374 & 0.6292 \\
HAtten-RR$^{\star}$        & 0.0001 & 0.0007 & 0.0001 & 0.0002 & 0.0003 & 0.0011 & 0.0027 & 0.0007 & 0.0016 \\
KTR$^{\star}$              & 0.0039 & 0.0142 & 0.0060 & 0.0071 & 0.0078 & 0.0172 & 0.0253 & 0.0271 & 0.0391 \\
AR-GNN$^{\star}$           & 0.0041 & 0.0140 & 0.0049 & 0.0067 & 0.0096 & 0.0284 & 0.0556 & 0.0361 & 0.0581 \\
LitFM$^{\star}$            & 0.0386 & 0.1226 & 0.0577 & 0.0700 & 0.0783 & 0.1626 & 0.2242 & 0.2501 & 0.3375 \\
\bottomrule
\end{tabular}%
\end{table*}

Table~\ref{tab:type1_results}, Table~\ref{tab:type2_results}, and Table~\ref{tab:type3_results} report retrieval performance for all baselines under the title-and-abstract input setting.
Unlike prior work on citation recommendation, which typically uses a single binary relevance label, our benchmark defines must-cite status through three ground truth types:
\textbf{Type~I} (experimental baseline, binary),
\textbf{Type~II} (core relevance, thresholded at score $\geq 4$), and
\textbf{Type~III} (intra-paper mention frequency, thresholded at $\geq 3$ mentions).
We report results under each ground truth independently to reveal which retrieval signals align best with each facet of must-cite status. All methods retrieve from the 67,761 candidate pool given only the query paper's
title and abstract as input.

\subsection{Cross-Type Summary.}
Table~\ref{tab:summary_all} provides a compact comparison of all baselines across the three ground truth types,
using Recall@100 as the representative metric. This view highlights which retrieval paradigms
are best aligned with each facet of must-cite status.

\begin{table}[ht]
\centering
\caption{Summary: Recall@100 across all baselines and ground truth types.}
\label{tab:summary_all}
\footnotesize
\begin{tabular}{l l c c c}
\toprule
\textbf{Type} & \textbf{Method} & \textbf{Type~I} & \textbf{Type~II} & \textbf{Type~III} \\
\midrule
Sparse
    & BM25              & 0.3006 & 0.2922 & 0.4139 \\
\midrule
\multirow{10}{*}{Dense}
    & SciBERT            & 0.0264 & 0.0270 & 0.0361 \\
    & SciBERT-NTX        & \textbf{0.3832} & \textbf{0.4102} & \textbf{0.4932} \\
    & SPECTER            & 0.2895 & 0.2785 & 0.3727 \\
    & SPECTER2          & 0.3258 & 0.3216 & 0.4221 \\
    & SciNCL             & 0.3223 & 0.3169 & 0.4163 \\
    & ColBERT            & 0.2497 & 0.2486 & 0.3317 \\
    & ColBERT-NTX        & 0.3416 & 0.3543 & 0.4579 \\
    & HAtten-RR          & 0.0031 & 0.0026 & 0.0027 \\
\midrule
\multirow{4}{*}{Graph}
    & KTR                & 0.0172     & 0.0193     & 0.0253     \\
    & AR-GNN             & 0.0573 & 0.0451 & 0.0556 \\
    & LitFM              & 0.2220 & 0.2520 & 0.2242 \\
\bottomrule
\end{tabular}%

\end{table}

\subsection{Analysis.}
\label{sec:analysis}

No method exceeds 50\% Recall@100 on any ground truth type: even the strongest baseline, SciBERT-NTX, leaves more than half of must-cite papers unrecovered across all three tiers. 
SciBERT-NTX, achieves Recall@100 of 0.383, 0.410, and 0.493 on Types~I, II, and III respectively (Tables~\ref{tab:type1_results}--\ref{tab:type3_results}),  confirming that must-cite retrieval is a strictly harder task than general citation recommendation and leaving substantial headroom for future methods.

\paragraph{Contrastive fine-tuning is the dominant signal, not architectural novelty.}
The two best methods on every ground truth type are both produced by applying NT-Xent contrastive loss to an off-the-shelf encoder: SciBERT-NTX on top of frozen SciBERT, and ColBERT-NTX on top of vanilla ColBERT. SciBERT-NTX improves over its base by over 13$\times$ in Recall@100; ColBERT-NTX gains a modest ${\sim}1.4\times$ over vanilla ColBERT. In contrast, citation-informed pre-training (SPECTER, SPECTER2, SciNCL) yields strong but not leading performance, and vanilla SciBERT collapses to near-zero on all metrics---suggesting that generic scientific-text pretraining provides little direct signal for must-cite retrieval without task-specific supervision.

\paragraph{BM25 remains surprisingly competitive on sparse-label ground truths.}
BM25 is the best method on most ranking metrics (MAP, MRR, nDCG@$K$, HR@$K$) for Type~I and Type~III, though it trails SciBERT-NTX on Recall@$K$ for every ground truth type. Type~I (experimental baselines) and Type~III (frequently mentioned papers) both tend to share distinctive surface vocabulary with the query---dataset names, method names, shared task terminology---which BM25 exploits directly. Type~II (core relevance), by contrast, rewards semantic proximity that keyword overlap cannot capture, and is where SciBERT-NTX pulls clearly ahead across nearly all metrics. The gap between sparse and dense methods therefore depends heavily on which facet of must-cite status is being measured, a distinction that a single-label benchmark would obscure.

\paragraph{Graph-based methods underperform text-only baselines in the cold-start setting.}
AR-GNN and LitFM, both of which rely on citation-graph structure at inference time, achieve Recall@100 roughly 6--8$\times$ and 1.6--2.2$\times$ lower than SciBERT-NTX, respectively. The 2025 query papers have no outbound citations in the training pool, so the structural neighbourhood these models exploit is unavailable at test time---a property of the must-cite auditing scenario rather than a general indictment of graph-based retrieval. HAtten-RR collapses to near-zero across all settings, consistent with its design for local, context-span-based citation recommendation rather than global title/abstract retrieval.

\paragraph{Type~II captures signal the other two tiers miss.}
SciBERT-NTX's Recall@100 on Type~II (0.410) exceeds its Type~I score (0.383), as does ColBERT-NTX, while SPECTER, SPECTER2, SciNCL, and vanilla ColBERT show roughly equal performance across both types. Type~I relies on explicit comparison language in the experiments section, which some papers genuinely lack even for papers that were core to the work; Type~III relies on repeat mentions, which long baseline-comparison tables can inflate mechanically. Type~II---the 1--5 core-relevance score---picks up papers that are central to the query's method or dataset without requiring either trigger, and the retrieval numbers are correspondingly higher. This supports the benchmark's three-tier design: the three ground truths are not interchangeable and each captures a distinct aspect of must-cite status.
\section{Discussion and Conclusion.}
\label{sec:conclusion}

\paragraph{Reusability. }
MasterSet's reusability rests on two design decisions independent of any specific annotation run. 
The first is \emph{data reliability}: by sourcing papers exclusively from official proceedings/websites rather than aggregator APIs, we obtain exact, authoritative paper counts free of preprint conflation---a property that no existing large-scale AI/ML corpus provides. Labels derived from a noisy candidate pool conflate collection artifacts with genuine citation relationships; the official-proceedings foundation eliminates this confound.
The second is \emph{modularity}: the three-tier scheme, the LLM judge
interface, and the threshold definitions are separable components,
allowing other communities to redefine must-cite for their own domain
or institutional criteria while reusing the collection and extraction
infrastructure wholesale.

\paragraph{Conclusion.}
Concretely, we built a benchmark of 153,373 papers from 15 official proceedings/websites, annotated 2,005,387 citation instances with a three-tier scheme, and showed that even the strongest baseline recovers fewer than half of must-cite papers in the top 100 from a 67,761-paper pool. 
To support reproducibility and future research, we release all venue-specific crawlers, paper metadata, data processing pipelines, annotation code, and baseline implementations.
Must-cite retrieval is a strictly open problem; future work should pursue architectures that exploit all three annotation tiers jointly, extend the query set beyond the three core ML venues, and ultimately integrate must-cite auditing into manuscript preparation workflows.

\newpage
\bibliographystyle{siamplain}
\bibliography{references}
\newpage
\appendix
\appendix
\section*{\centering Appendix}
\addcontentsline{toc}{section}{Appendix}

\section{Semantic Scholar Discrepancy Statistics}
\label{app:s2orc}

To substantiate our claim that aggregator services return inconsistent paper counts, we queried the Semantic Scholar API for every conference-year pair in our AAAI, NeurIPS, ICLR, and ICML collections and compared the results against our official-proceedings ground truth. A \emph{spurious} entry is a record returned by the API that does not correspond to a main-conference paper in the official proceedings (e.g., workshop papers, symposium volumes, or proceedings-level records). A \emph{missing} entry is a verified main-conference paper absent from the API response.

Table~\ref{tab:s2orc} summarises the results. Discrepancies are present in every venue and every year tested, and grow substantially with conference size and recency. Note that the extreme missing counts for 2024--2025 reflect indexing lag (papers not yet ingested) rather than structural errors in the API; all other years reflect steady-state retrieval behaviour.

\begin{table*}[h]
\centering
\caption{Semantic Scholar API discrepancies relative to official proceedings,
by venue. Spurious: records returned by API not present in official
proceedings. Missing: official papers absent from API response. Worst-year
figures exclude 2024--2025 to avoid conflating indexing lag with structural
errors.}
\label{tab:s2orc}
\footnotesize
\begin{tabular}{lrrrll}
\toprule
\textbf{Venue} & \textbf{Years} & \textbf{Total} & \textbf{Total} &
\textbf{Worst year} & \textbf{Worst year} \\
 & \textbf{tested} & \textbf{spurious} & \textbf{missing} &
\textbf{(spurious)} & \textbf{(missing)} \\
\midrule
AAAI    & 2010--2025 & 7{,}861 & 6{,}715 &
    2023 ($+$1{,}231) & 2023 ($-$806) \\
NeurIPS & 2000--2024 & 2{,}622 & 2{,}105 &
    2023 ($+$381)   & 2023 ($-$351) \\
ICLR    & 2013--2025 & 6{,}677 & 6{,}343 &
    2024 ($+$1{,}515) & 2024 ($-$1{,}295) \\
ICML    & 2013--2025 & 2{,}662 & 5{,}890 &
    2023 ($+$485)   & 2023 ($-$427) \\
\midrule
\textbf{Total} & & \textbf{19{,}822} & \textbf{21{,}053} & & \\
\bottomrule
\end{tabular}
\end{table*}

\paragraph{Examples of spurious entries.}
Inspection of the spurious records reveals three recurring patterns: (i) proceedings-level records (e.g., ``Proceedings of the Twenty-Fourth AAAI Conference on Artificial Intelligence'' returned as a paper); (ii) workshop and symposium papers whose metadata shares the same venue string as the main conference; and (iii) duplicate records arising from preprint--published version conflation. These patterns motivate our decision to collect papers exclusively from official proceedings pages rather than via aggregator APIs.

\section{Nougat Failure Mode Details.}
\label{app:failures}

\paragraph{Case~I: Reference reordering.}
Nougat occasionally renumbers references in the order it first encounters them in the body text, rather than preserving the original manuscript numbering. For example, a paper whose reference section lists entry~[5] as Smith et al.\ and entry~[13] as Jones et al.\ may be rendered with those entries reassigned to~[1] and~[2]. As a consequence, an in-text citation of \texttt{[13]} is mapped by our extractor to the wrong reference entry, silently introducing an incorrect citation--context association. This failure mode is invisible to heuristics that operate on the Nougat output alone, because the renumbered references and in-text keys are internally consistent within the corrupted file.

\paragraph{Case~II: OCR misreadings.}
Nougat is fundamentally an OCR model and occasionally misreads individual digits in numeric citation keys. A common example is the confusion of \texttt{1} and \texttt{l} or \texttt{0} and \texttt{O}, producing erroneous keys such as \texttt{[l0, 33, 22]} instead of \texttt{[10, 33, 22]}. These errors cause citation look-ups to fail silently: the key is not found in the reference list, and the corresponding context is lost rather than incorrectly attributed.

\paragraph{Case~III: Alpha-tag / numeric mismatch.}
A subset of papers typeset their reference lists with alpha-tags (e.g., \texttt{[DKS17]}, \texttt{[GDDM14]}, \texttt{[KPR+17]}) but Nougat converts all in-text citations to sequential numeric keys, because its training data predominantly features numeric citation styles. The reference list and the body of the document then use incompatible key systems, making it impossible to link any in-text citation to its reference entry. We detect this mismatch automatically: if more than 30\% of reference entries begin with an alpha-tag and the body contains predominantly numeric citations \texttt{[\textit{n}]}, the file is flagged and excluded from Nougat processing entirely.

Additional minor failure modes---most notably missing sections, where Nougat omits an entire section (such as the Introduction) from its output---are comparatively rare and do not affect the reference-mapping step.

\section{Output Schema Specification.}
\label{app:schema}

Regardless of the extraction path, every paper produces a single JSON file containing a list of reference objects. Each object stores: the internal reference identifier (\texttt{target}), the reference title (\texttt{title}), the publication year (\texttt{year}), the in-text cite string as it appears in the body (\texttt{cite}), and a list of citation contexts (\texttt{contexts}). Each context record stores the inferred section label (\texttt{section}) and the three-sentence window surrounding the citation (\texttt{context}).

Two post-processing filters are applied uniformly across both paths:
\begin{itemize}
    \item \textbf{Year filter.} Reference entries for which a four-digit year cannot be parsed are discarded. Such entries are overwhelmingly figures, captions, or footnotes that Nougat misplaces into the reference section.
  \item \textbf{Context deduplication.} If the same three-sentence window is attributed to more than one co-cited reference (as occurs with group citations such as \texttt{[3, 7, 14]}), it is retained for each individual reference but a seen-set prevents exact duplicates within a single reference's context list.
\end{itemize}

\section{Pipeline Coverage Statistics.}
\label{app:pipeline}

Table~\ref{tab:pipeline} reports the distribution of papers across the two extraction tiers. Of the \textbf{153,373} papers in the collection, \textbf{143,970} (\textbf{93.87}\%) are successfully processed by Nougat; and from the remaining, \textbf{9,348} (\textbf{6.09}\%) are processed by GROBID, covering papers that triggered reference reordering (Cases~I--II, concentrated in IJCAI, NeurIPS, and UAI) or alpha-tag mismatch (Case~III, scattered across all venues). The residual 0.04\% (55 papers) are scanned image-based PDFs for which both Nougat (already flagged via the alpha-tag heuristic) and GROBID---which requires an extractable text stream---fail to produce usable output.

\begin{table}[ht]
\centering
\caption{Distribution of papers across the two-tier extraction pipeline.}
\label{tab:pipeline}
\footnotesize
\begin{tabular}{lrr}
\toprule
\textbf{Extraction path} & \textbf{Papers} & \textbf{Share} \\
\midrule
Nougat (primary)  & 143,970 & 93.87\% \\
GROBID (fallback) & 9,348 & 6.09\% \\
\midrule
\textbf{Total processed} & \textbf{153,318} & \textbf{99.96\%} \\
\bottomrule
\end{tabular}
\end{table}

\section{Annotation Details.}
\label{app:annotation}

The following three subsections provide complete documentation for the annotation pipeline: the criteria used to define must-cite status (Section~\ref{app:tier_criteria}), the LLM prompts used to operationalise Types~I and~II at scale (Section~\ref{app:prompts}), and the human validation protocol used to verify label quality (Section~\ref{app:protocol}). The criteria in Section~\ref{app:tier_criteria} are embedded verbatim in the prompts in Section~\ref{app:prompts}; both are reproduced here for completeness.

\subsection{Three-Tier Annotation Criteria.}
\label{app:tier_criteria}

\subsubsection*{Type~I: Experimental Baseline}

We assign $y^{\mathrm{I}} = 1$ if and only if:
\begin{itemize}[leftmargin=*]
  \item the cited work is directly compared against the proposed method in the \emph{current paper};
  \item the cited work is explicitly used as a baseline in experiments or results;
  \item the cited work is the source of a main benchmark, dataset, or evaluation task used in the \emph{current paper}; or
  \item the cited work defines a core task, problem setting, or method that the \emph{current paper} directly builds on, evaluates on, or centres around.
\end{itemize}
A citation does \emph{not} qualify if it appears only in Related Work or Introduction, describes a method the current paper uses but does not compete against, or if the comparison language refers to what other papers did rather than the current paper.

\subsubsection*{Type~II: Core Relevance Rubric}

\begin{enumerate}[leftmargin=*, label=\arabic*.]
  \item \textbf{General background}, unrelated to this paper's task or method. \emph{Example: a general NLP survey cited in a citation classification paper.}
  \item \textbf{Tangentially related}; shares a broad area but not this specific task or method. \emph{Example: a general BERT paper cited in a scientific document retrieval paper where BERT is not the proposed method.}
  \item \textbf{Conceptually related} but neither the same task nor a directly extended method. \emph{Example: a text classification paper cited in a citation intent classification paper.}
  \item \textbf{Substantially relevant}: same task OR a key method component is derived from it. \emph{Example: SciBERT cited in a paper that fine-tunes SciBERT for citation classification.}
  \item \textbf{Core relevance}: same task AND method is directly extended, OR defines the primary dataset or benchmark used. \emph{Example: the ACL-ARC dataset paper cited in a paper that evaluates on ACL-ARC.}
\end{enumerate}

Two hard rules apply regardless of context: a citation to the primary dataset or benchmark receives at least~4; a citation to a directly extended method receives at least~4; a citation satisfying both conditions receives~5.

\subsubsection*{Type~III: Mention Frequency Threshold}

\begin{figure*}[ht]
\centering
\includegraphics[width=\textwidth]{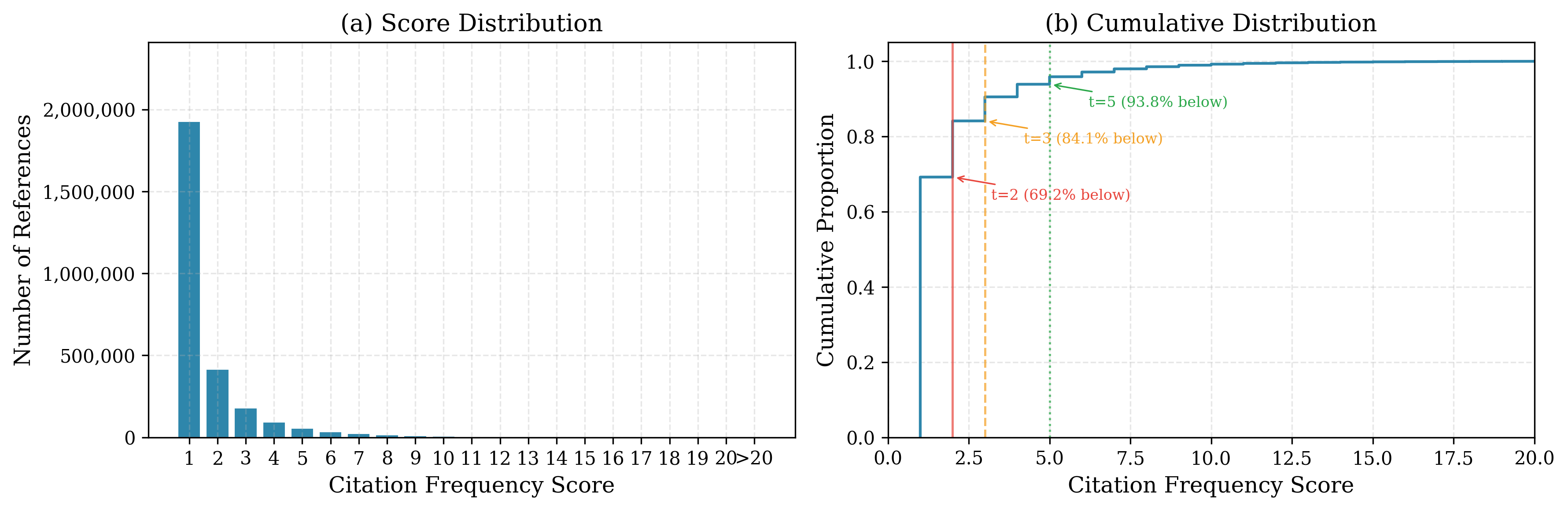}
\caption{Empirical distribution (left) and cumulative distribution (right) of intra-paper mention frequency across all annotated citation instances. Candidate thresholds $t \in \{2, 3, 5\}$ are overlaid on the CDF with the fraction of instances below each. $N = 3$ is selected as the must-cite threshold: the $N = 2 \to N = 3$ transition cleanly separates incidental (1--2 mention) from sustained engagement, while larger thresholds admit noise from heavy-citation outliers.}
\label{fig:type3_threshold}
\end{figure*}

The threshold $N$ is selected by inspecting the empirical distribution of within-paper mention counts across all annotated citation instances (Figure~\ref{fig:type3_threshold}). The distribution is heavily right-skewed: 69.2\% of references are mentioned exactly once, and 84.1\% are mentioned one or two times. References mentioned three or more times --- the candidates for a Type~III must-cite flag --- comprise only 15.9\% of instances.

We select $N = \textbf{3}$ on two grounds. First, references mentioned once or twice are overwhelmingly background citations: their single or paired appearance typically signals acknowledgement of prior work rather than sustained engagement. Setting $N = 2$ would admit 30.8\% of instances as must-cite, diluting the intended ``core dependency'' semantics of the tier. Second, $N = 3$ is the smallest threshold at which a reference is likely to have appeared in more than one section of the paper --- the operational definition of sustained engagement we intend the tier to capture. Moving to $N = 5$ would tighten precision further (only 6.2\% of instances qualify) but at the cost of excluding many references that are genuinely engaged with across motivation, methodology, and experiments without being mentioned five or more times; the 84.1\% $\to$ 93.8\% jump between $N = 3$ and $N = 5$ in the CDF is dominated by papers that cite a few core works very heavily, which is a noisier signal than the 1--2-to-3 transition. $N = 3$ therefore provides the cleanest separation between incidental and sustained citation, and we adopt it as the Type~III threshold for the benchmark.

\subsection{LLM Annotation Prompts.}
\label{app:prompts}

We reproduce the final Type~I and Type~II system prompts used for LLM annotation. Each prompt is submitted to Gemini 2.5 Flash via the API's native \texttt{system\_instruction} and \texttt{contents} fields; no model-specific special tokens are hard-coded in the prompt string. The model is run in greedy decoding mode with a maximum of 4 new tokens.

\subsubsection{Type~I Prompt: Experimental Baseline Detection.}
\label{app:prompt_type1}

\begin{small}
\begin{spverbatim}
[SYSTEM]
You are an expert research assistant in AI and machine learning.

Your task is to determine whether a cited work is either:
1. a baseline or direct comparison method in the current paper, OR
2. a must-cite core reference for the current paper's main task,
   benchmark, dataset, or method.

A citation qualifies (answer 1) IF ANY of the following are true:
- The cited work is directly compared against the proposed method
  in the CURRENT PAPER
- The cited work is explicitly used as a baseline in experiments
  or results
- The cited work is the source of a main benchmark, dataset, or
  evaluation task used in the CURRENT PAPER
- The cited work defines a core task, problem setting, or method
  that the CURRENT PAPER directly builds on, evaluates on, or
  centers around

Strong evidence for answer 1 includes:
- Language such as "compared to", "outperforms", "evaluated
  against", "our method vs.", "surpasses", "baseline" in
  Experiments or Results
- Statements in Experimental Setup or Experiments showing that
  the paper evaluates on a benchmark/dataset introduced by the
  cited work
- Statements showing the cited work defines the main
  task/problem central to the current paper

A citation does NOT qualify (answer 0) if:
- It is cited only for background, motivation, or general context
- It appears only as loosely related prior work
- It describes a method or resource that is mentioned but not
  central to the paper's main experiments, task, or contribution
- The comparison language refers to what OTHER papers did, not
  the CURRENT PAPER
- It is a peripheral citation rather than a core benchmark, task,
  dataset, or comparison target

Important:
- The decision must be based on the CURRENT PAPER, not on what
  the cited paper itself did
- A citation can qualify even if it is not a baseline, as long as
  it is a must-cite core reference for the paper's main benchmark,
  dataset, task, or method
- Citation contexts may be sentence fragments, so use the section
  label carefully

[EXAMPLE -- Answer: 1]
1. Section: "4 Experiments"
   Context: "Our method achieves 84.2 F1, outperforming
   [CITATION] (79.1) on SciERC."
-> Direct baseline/comparison in results. Answer: 1

[EXAMPLE -- Answer: 1]
1. Section: "4 Experimental Setup"
   Context: "We evaluate our model on StrategyQA [CITATION],
   a benchmark for implicit reasoning."
-> The cited work is the source of a main benchmark used in
   the paper. This is a must-cite core reference. Answer: 1

[EXAMPLE -- Answer: 0]
1. Section: "2 Related Work"
   Context: "Previous work such as [CITATION] explored
   prompt-based methods for NLP."
-> Background reference only. Answer: 0

Respond with ONLY a single digit: 0 or 1.
No explanation. No punctuation.

[USER]
Paper Title: {title}

Citation: "cited_as": "{cited_as}", "title": "{ref_title}"

Citation Contexts:
{formatted_contexts}

Is this citation used as a baseline or direct comparison,
or is it a must-cite core reference in this paper?
Answer:
\end{spverbatim}
\end{small}

\subsubsection{Type~II Prompt: Core Relevance Scoring.}
\label{app:prompt_type2}

\begin{small}
\begin{spverbatim}
[SYSTEM]
You are an expert research assistant in AI and machine learning.
Score the relevance of a citation to the CORE TASK AND METHOD
of a paper (1-5 scale).

1  General background, unrelated to this paper's task or method.
   Example: A general NLP survey cited in a citation
   classification paper.

2  Tangentially related; shares a broad area but not this
   specific task or method.
   Example: A general BERT paper cited in a scientific document
   retrieval paper where BERT is not the proposed method.

3  Conceptually related but neither the same task nor a directly
   extended method.
   Example: A text classification paper cited in a citation
   intent classification paper.

4  Substantially relevant: same task OR a key method component
   is derived from it.
   Example: SciBERT cited in a paper that fine-tunes SciBERT
   for citation classification.

5  Core relevance: same task AND method is directly extended,
   OR defines the primary dataset/benchmark used.
   Example: The ACL-ARC dataset paper cited in a paper that
   evaluates on ACL-ARC.

Rules:
- Primary dataset/benchmark citation -> at least 4
- Method directly extended -> at least 4
- Both task AND method shared -> 5

Note: Citation contexts may be sentence fragments due to
extraction. Use the section label and available context to
make your best judgment.

Respond with ONLY a single digit: 1, 2, 3, 4, or 5.
No explanation. No punctuation.

[USER]
Paper Title: {title}

Citation: "cited_as": "{cited_as}", "title": "{ref_title}"

Citation Contexts:
{formatted_contexts}

How relevant is this citation to the paper's core task
and method?
Score:
\end{spverbatim}
\end{small}

\subsubsection{Prompt Development: Initial vs.\ Revised.}
\label{app:prompt_changes}

Table~\ref{tab:prompt_changes} summarises the key design changes made during prompt development. The initial prompts used LLaMA-style special tokens (e.g., \texttt{<|begin\_of\_text|>}, \texttt{<|start\_header\_id|>}) hard-coded in the prompt string, which are not used by Gemini 2.5 Flash and were replaced with model-native API formatting. The initial Type~I prompt recognised only baselines and direct comparisons; the revised prompt adds a second qualifying category---must-cite core references for the paper's main benchmark, dataset, task, or method---with four explicit positive conditions and a dedicated few-shot example for the benchmark-source case. The ``any context'' rule in the initial Type~I prompt caused false positives when a citation context described a comparison made by another paper; the revised scope guard, promoted to an explicit \texttt{Important} block, restricts the qualifying condition to the current paper. The initial Type~II rubric provided no concrete examples for levels 3, 4, and 5, causing the model to collapse scores toward the centre; the revised rubric anchors each level with one representative example.

\begin{table*}[ht]
\centering
\caption{ Key changes between the initial (zero-shot) and final (revised) annotation prompts. }
\label{tab:prompt_changes}
\small
\begin{tabular}{p{2.8cm}p{4.5cm}p{5.5cm}}
\toprule
\textbf{Aspect} & \textbf{Initial} & \textbf{Revised} \\
\midrule
Stage~1 scope &
  Baselines / direct comparisons only &
  Two categories: baseline OR must-cite core reference
  (benchmark, dataset, task, method) \\
\addlinespace
Positive conditions &
  Single composite rule triggered by keyword match &
  Four specific conditions, any of which is sufficient \\
\addlinespace
Evidence guidance &
  Trigger-word list &
  Trigger words plus section-location cues
  (Experiments, Experimental Setup) \\
\addlinespace
Current-paper guard &
  ``If any context indicates\ldots'' &
  Promoted to an \texttt{Important} block; decision is
  about the current paper, not the cited one \\
\addlinespace
Few-shot examples &
  None &
  Three anchored examples in Type~I (positive comparison,
  positive benchmark-source, negative background);
  one per level in Type~II \\
\addlinespace
Rubric anchoring &
  Abstract descriptions only &
  One concrete example per score level, plus a
  \texttt{Rules} block fixing the floor for benchmark
  and method citations \\
\addlinespace
Fragment handling &
  Not mentioned &
  Explicit note to use the section label when contexts
  are truncated \\
\addlinespace
Output instruction &
  Prose: ``do not provide reasoning'' &
  Structural: ``ONLY a single digit\ldots No explanation.
  No punctuation.'' \\
\addlinespace
Token format &
  LLaMA-style tokens hard-coded in prompt string &
  Clean \texttt{[SYSTEM]}/\texttt{[USER]} blocks sent
  via the Gemini API \\
\addlinespace
Max new tokens &
  96 &
  4 \\
\bottomrule
\end{tabular}
\end{table*}
\subsection{Annotation Cost and Model Selection.}
\label{app:cost}

Annotating all 2,005,387 citation instances for Types~I and~II via the Gemini~2.5 Flash API incurred a total cost of \textbf{\$619.41}, or approximately \$0.000309 per instance.

\paragraph{Why Gemini~2.5 Flash.}
Types~I and~II are fundamentally classification tasks---a binary judgment and a five-point ordinal score, respectively---that require careful reading of citation contexts but not extended chain-of-thought reasoning. A capable but economical model is therefore sufficient; the frontier reasoning capacity of larger models adds latency and cost without improving label reliability for these well-structured, few-shot-anchored prompts. Gemini~2.5 Flash matches the quality bar established by our human validation study (Table~\ref{tab:agreement}) while keeping the per-instance cost low enough to make annotation at dataset scale tractable.

\paragraph{Input-heavy, output-light cost structure.}
Each annotation call submits a query paper title, the cited paper's bibliographic entry, and all extracted citation contexts with section labels---a long input---but requires only a single digit in response (maximum four new tokens, Section~\ref{sec:llm_pipeline}). Most commercial LLM APIs charge output tokens at a premium over input tokens; the extreme asymmetry of our prompts therefore exploits the pricing structure efficiently, keeping total cost low despite the volume of input text processed.

\subsection{Human Validation Annotation Protocol.}
\label{app:protocol}

The 510 annotation instances are distributed across annotators such that every instance is independently labelled by between two and six people, with no discussion between annotators. Each instance presents: the query paper title, the cited paper title and year, and all extracted citation contexts with their section labels.

\textbf{Question~1 (Type~I)}: Is the cited paper used as a direct comparison baseline in the experiments of \emph{this} paper? Response: $1 = \text{Yes}$, $0 = \text{No}$.

\textbf{Question~2 (Type~II)}: How relevant is the cited paper to the core task and method of the query paper? Response: integer from 1 to 5, using the scale defined in Section~\ref{sec:tiers}.

Annotators are given the following instructions:
\begin{itemize}[leftmargin=*]
  \item For Q1, judge only the \emph{current} paper's experiments, not what other papers have done with the cited work.
  \item Citation contexts may be sentence fragments due to PDF extraction; use the section label as an additional signal.
  \item Q1 and Q2 are independent: a paper can be a baseline (Q1=1) without being highly relevant to the core method (Q2$\leq$3), and vice versa.
\end{itemize}

\subsection{Inter-Annotator Agreement Supporting Statistics.}
\label{app:iaa_details}

Table~\ref{tab:iaa_full} reports the full set of agreement statistics on the 510-instance validation set. Krippendorff's~$\alpha$ is the primary metric because it supports the variable number of raters per item (2--6) natively. For Type~I we additionally report the proportion of instances on which every rater agreed, and the distribution of pairwise Cohen's~$\kappa$ over the $\binom{6}{2} = 15$ annotator pairs, computed on each pair's co-rated subset. For Type~II we report $\alpha$ under both ordinal and interval weights, the proportion of instances where the rater spread ($\max - \min$) is at most~1, and the distribution of pairwise quadratic-weighted Cohen's~$\kappa$. Following the Landis \& Koch interpretation~\cite{landis1977measurement}, both tasks fall in the \emph{moderate} range (0.41--0.60), which is consistent with prior citation-annotation datasets on subjective sub-labels~\cite{cohan2019structural, jurgens2018measuring}.

\begin{table*}[ht]
\centering
\caption{Full inter-annotator agreement statistics on the 510-instance human validation set (2--6 raters per instance).}
\label{tab:iaa_full}
\small
\begin{tabular}{lcc}
\toprule
\textbf{Metric} & \textbf{Type~I} & \textbf{Type~II} \\
\midrule
Krippendorff's~$\alpha$ (nominal)              & \textbf{0.461} & --- \\
Krippendorff's~$\alpha$ (ordinal)              & ---   & \textbf{0.492} \\
Krippendorff's~$\alpha$ (interval)             & ---   & 0.490 \\
\midrule
\% exact agreement (all raters on an item)     & 69.0\% & 27.1\% \\
\% within $\pm 1$ (all raters on an item)      & ---    & 63.7\% \\
\midrule
Pairwise Cohen's~$\kappa$ (mean / min--max)    & 0.480 / 0.26--0.64 & --- \\
Pairwise weighted~$\kappa$ (mean / min--max)   & --- & 0.478 / 0.15--0.78 \\
\bottomrule
\end{tabular}
\end{table*}

\section{Baseline Descriptions.}
\label{app:baselines}

\subsection*{BM25~\cite{robertson1995okapi}}
BM25 ranks candidates by keyword matching between query and candidate text using term frequency, inverse document frequency, and document length normalisation. We index the same textual signals as the semantic baselines (title + abstract) and retrieve the top candidates by BM25 score.

\subsection*{SciBERT~\cite{beltagy2019scibert}}
SciBERT is a BERT-based encoder pre-trained on scientific text (biomedical and computer science). We use pre-trained weights to generate fixed document-level embeddings and retrieve via cosine similarity. We additionally fine-tune SciBERT with NT-Xent contrastive loss (SciBERT-NTX) to assess the benefit of task-specific fine-tuning objectives.

\subsection*{SPECTER~\cite{cohan2020specter}}
SPECTER pre-trains SciBERT-based embeddings using citation-informed triplet supervision: cited papers are pulled closer to the query than uncited negatives. Inference requires only title and abstract.

\subsection*{SPECTER2~\cite{Singh2022SciRepEvalAM}}
SPECTER2 extends SPECTER with improved multi-task training objectives, learning representations suitable for multiple scientific NLP tasks simultaneously.

\subsection*{SciNCL~\cite{Ostendorff2022scincl}}
SciNCL replaces SPECTER's triplet loss with neighbourhood contrastive learning, drawing on the citation graph neighbourhood to define positive and negative pairs.

\subsection*{ColBERT~\cite{khattab2020colbert}}
ColBERT uses late interaction: query and document token embeddings are pre-computed and stored; relevance is scored at query time via MaxSim operations over all query--document token pairs. We evaluate vanilla ColBERT and a variant fine-tuned with NT-Xent contrastive loss (ColBERT-NTX).

\subsection*{HAtten-RR~\cite{gu2022local}}
HAtten-RR is a two-stage pipeline. A hierarchical attention network prefetches top candidates via cosine similarity; a SciBERT-based reranker then scores them with a feed-forward network. Both stages are trained with triplet margin loss.

\subsection*{KTR~\cite{wu2024supporting}}
KTR combines BERT-based text encoding with GCNs and a GRU to model topic-level reasoning paths. A Content-Structure Alignment module uses contrastive learning to simulate citation-graph topology from text alone, enabling cold-start use. The final ranking score combines global semantic similarity with explicit reasoning paths over a common-sense knowledge graph.

\subsection*{AR-GNN~\cite{kammari2024relevant}}
AR-GNN learns from heterogeneous bibliographic graphs with paper, author, venue, and keyword nodes. Three variants---AR-GCN, AR-GAT, and AR-GAE---use different GNN architectures. Training uses a custom loss combining Euclidean distance and mean similarity to map related papers to nearby embeddings.

\subsection*{LitFM~\cite{zhang2025litfm}}
LitFM combines a BERT text encoder with a GNN layer that aggregates citation-graph neighbourhood information. An MLP reconstructs pseudo-query embeddings for ambiguous inputs. Training uses InfoNCE loss with L1 regularisation.

Table~\ref{tab:baseline_summary} summarises architectures and training objectives.

\begin{table*}[t]
\centering
\caption{Summary of baseline methods.}
\label{tab:baseline_summary}
\footnotesize
\begin{tabular}{llll}
\toprule
\textbf{Baseline} & \textbf{Architecture} &
\textbf{Training Objective} & \textbf{Inference} \\
\midrule
BM25            & Sparse lexical retriever      & None (unsupervised)         & BM25 score ranking \\
SciBERT         & Transformer (BERT-Base)       & MLM (pretrained)            & Cosine similarity \\
SciBERT-NTX     & Transformer (BERT-Base)       & NT-Xent contrastive         & Cosine similarity \\
SPECTER         & Transformer (SciBERT)         & Triplet margin loss         & Cosine similarity \\
SPECTER2        & Transformer (SciBERT)         & Multi-task triplet          & Cosine similarity \\
SciNCL          & Transformer (SciBERT)         & Neighbourhood contrastive   & Cosine similarity \\
ColBERT         & Late-interaction Transformer  & Pairwise softmax            & MaxSim \\
ColBERT-NTX     & Late-interaction Transformer  & NT-Xent contrastive         & MaxSim \\
HAtten-RR       & Hierarchical Attn + SciBERT   & Triplet margin loss         & Cosine (prefetch) + MLP (rerank) \\
KTR             & BERT + GCN + GRU              & BCE + contrastive           & Cosine + reasoning score \\
AR-GNN          & Heterogeneous GNN             & Euclidean + similarity loss & Cosine similarity \\
LitFM           & BERT + GNN + MLP              & InfoNCE + L1 regularisation & Cosine similarity \\
\bottomrule
\end{tabular}
\end{table*}

\section{Evaluation Metrics.}
\label{app:metrics}

\paragraph{MAP (Mean Average Precision).}
MAP measures ranking quality by averaging precision at each rank position where a relevant item appears, then averaging across queries. It rewards models that place must-cite papers at the top of the list.

\paragraph{MRR (Mean Reciprocal Rank).}
MRR measures how quickly the first relevant item appears:
\begin{equation}
  \mathrm{MRR} = \frac{1}{|Q|}\sum_{i=1}^{|Q|}\frac{1}{\mathrm{rank}_i}.
\end{equation}
A first hit at rank~1 scores~1; at rank~5 it scores~0.2.

\paragraph{nDCG@$K$ (Normalised Discounted Cumulative Gain).}
nDCG@$K$ evaluates ordering quality in the top~$K$ with logarithmic position discounting:
\begin{equation}
  \mathrm{nDCG}@K = \frac{\mathrm{DCG}@K}{\mathrm{IDCG}@K}, \quad
  \mathrm{DCG}@K = \sum_{i=1}^{K}\frac{rel_i}{\log_2(i+1)},
\end{equation}
where $\mathrm{IDCG}@K$ is the gain of the ideal ranking.

\paragraph{Recall@$K$.}
Recall@$K$ measures what fraction of all must-cite papers appear in the top~$K$ retrieved results. High recall is the primary objective for must-cite retrieval, where missing a foundational paper is costly.

\paragraph{HR@$K$ (Hit Ratio).}
HR@$K$ is a binary metric that equals~1 if the top~$K$ list contains at least one must-cite paper, and~0 otherwise:
\begin{equation}
  \mathrm{HR}@K = \frac{1}{|Q|}\sum_{i=1}^{|Q|}
  \mathbb{I}(\mathrm{hits}_i > 0).
\end{equation}

\section{Ethical Considerations.}
\label{app:ethics}

\subsection*{Data Collection and Copyright}

All papers are obtained from official conference websites providing publicly accessible content without authentication or subscription. Specifically: \texttt{aclanthology.org} licenses materials published since 2016 under CC~BY~4.0; \texttt{openreview.net} explicitly supports academic data access; \texttt{openaccess.thecvf.com} and \texttt{proceedings.mlr.press} are open-access platforms whose stated purpose is scholarly dissemination. No technical access controls, paywalls, or login barriers were circumvented. Venues whose proceedings require subscription access (e.g., KDD via ACM~DL, ICDM via IEEE~Xplore) were deliberately excluded. We release paper metadata (title, authors, abstract, venue, year), citation graphs, LLM-generated labels, and the collection pipeline source code; PDF files are not redistributed.

\subsection*{Venue Coverage and Representativeness}

According to Google Scholar Metrics h5-index rankings (2020--2024), the top-ranked AI and Computer Vision venues are CVPR (422), NeurIPS (309), ICLR (303), ICML (254), ECCV (238), ICCV (228), AAAI (212), IJCAI (133), and JMLR (106)~\cite{googlescholar2026}; all nine appear in our collection. ACL, EMNLP, and NAACL are the flagship NLP venues; AISTATS, UAI, and COLT cover probabilistic methods and learning theory.

Systematic coverage gaps include: (i)~subscription-access venues such as KDD, ICDM, and SIGIR; (ii)~workshops and co-located events; (iii)~arXiv preprints; and (iv)~non-English venues. Benchmark scores should be interpreted within the scope of the 15 covered venues; a method achieving high Recall@$K$ may still miss must-cite papers appearing exclusively in uncovered venues.

\subsection*{LLM-Generated Ground Truth}

LLMs may exhibit systematic biases---favouring English-speaking institutions, highly cited works, or papers stylistically similar to training data---that would propagate into benchmark labels. We mitigate this through few-shot anchoring and structured output constraints, and quantify residual error through the human validation study (Section~\ref{sec:validation}). Errors in citation context extraction also propagate to the LLM judge; pipeline provenance is recorded per paper to allow stricter downstream filtering.

\subsection*{Dual-Use Considerations}

Automated must-cite recommendation can help researchers discover overlooked prior work, but could also be used to identify citation patterns for superficial compliance with reviewer requests. We encourage deployment as an aid to scholarly diligence rather than a substitute for researcher judgement. Must-cite labels reflect community citation patterns; they do not constitute normative claims about what any paper ought to cite.

\end{document}